\documentclass[12pt]{article}
\usepackage{amssymb}
\begin{document}
 \author{
   Vladimir F.~Kovalev\footnote{
   Institute for Mathematical Modeling, RAS,
   Moscow, Russia, e-mail: kovalev@imamod.ru}
   \ and
   Dmitrij V.~Shirkov\footnote{
   Bogoliubov Laboratory for Theoretical Physics,
   Joint Institute for Nuclear Research, Dubna, Russia,
   \hspace*{0.5cm}
   e-mail: shirkovd@thsun1.jinr.ru}}
 \title{Functional self-similarity and renormalization group symmetry in
        mathematical physics}
 \date{}
 \maketitle
\begin{abstract}
The results from developing and applying the notions of functional
self-similarity and the Bogoliubov renormalization group to
boundary-value problems in mathematical physics during the last
decade are reviewed. The main achievement is the regular algorithm
for finding renormalization group--type symmetries using the
contemporary theory of Lie groups of transformations.
\end{abstract}

\section{Introduction}

The notion of functional self-similarity (FS) was introduced in
mathematical physics by one of the authors in the early
1980s~\cite{dv82} (see also \cite{dv84,dv88}). The basis of this
introduction is that solutions of a wide class of problems
analyzed by the renormalization group (RG) method are invariant
w.r.t.\ the group transformations that involve not only natural
independent variables of a problem but also parameters of boundary
conditions imposed at some ``reference" point. The RG
transformation then corresponds to reparameterizing a solution by
changing (shifting or rescaling) an independent (coordinate)
variable while simultaneously performing a functional
transformation of (boundary) characteristics of the selected
functions when passing to another reference point. The
corresponding ``transformation functions" are governed by group
functional equations.

More precisely, we consider the so-called renormalization
transformations (the Dyson transformations) in quantum field
theory (QFT), which constitute a continuous one-parameter group,
i.e., the Lie group of transformations (if we use a standard
mathematical language). This group was discovered~\cite{stp} and
used~\cite{bsh55} to analyze QFT singularities. We call this group
the QFT RG or the {\sl Bogoliubov} RG to distinguish it from the
approximate RG, which was introduced by Wilson~\cite{ken72} to
analyze critical phenomena in statistical physics problems.

The reasoning above pertains to the RG transformations considered
within the QFT approach. In the middle 1980s, the RG method became
widely applied to classical problems in mathematical physics; this
was initially connected with transforming and applying methods
developed for QFT and statistical physics. Such applications were
based, first, on the fact that physical problems manifest the FS
property, which permits segregating the characteristic variables
of a problem (independent and dynamic variables, parameters, and
boundary data) and finding transformations that preserve the
solution, and, second, on an advantageous mathematical description
that uses only differential and integral-differential equations to
describe the model. The RG approach is also favorable because the
corresponding RG-type transformations can be constructed using
regular methods, which are used to find symmetries in the group
analysis of differential equations (DE). Substantial progress has
been achieved in group analysis since the RG method first appeared
in theoretical physics several decades ago.

The transition to boundary-value problems in mathematical physics
enriched the mathematical content while preserving the main
property of the FS, i.e., the solution of a physical problem is
invariant w.r.t.\ a special class of transformations. The initial
notion of the FS transformations as point transformations of
independent variables was recently generalized to contact
transformations, transformations defined by formal series, etc.
Because the infinitesimal transformation approach is convenient
for the group analysis of mathematical models based on DEs, we can
also formulate the FS in terms of infinitesimal operators
determining the corresponding RG symmetry (RGS). The operators
then appear as a result of the standard RGS construction
procedure, and the FS condition arises in the course of this
procedure.

We review the evolution of the FS notion connected with the
implementation of the RGS method in mathematical physics and
describe the results obtained in the RGS framework. In Sec.~2, we
introduce both infinitesimal and finite RG transformations,
establish their connection with the mathematical physics notion of
powerlike self-similarity (automodelness), and introduce the
notion of the FS. We then construct RGSs in mathematical physics.
The FS property appears as an ingredient of this construction, and
we trace how the FS notion changes when passing from the
Bogoliubov RG to mathematical physics models.

In Sec.~3, we consider examples of the FS transformations that are
close to the QFT transformations because they are realized as
groups of one-parameter transformations w.r.t.\ some independent
variables. We present examples where the group extends and
generates FS transformations with operators constituting a
finite-dimensional algebra. Because an ODE or a system of such
equations governing the mathematical model discussed in this
section coincides with the Lie equations arising in the QFT
applications of the RG method, the formalism developed for ODEs
can also be applied to Lie equations.

In Sec.~4, we collect examples of FS conditions written either as
a first-order PDE or as higher-order differential relations. We
discuss a novel form of the FS transformations that can be
presented as formal infinite series, not as traditional algebraic
relations. The FS transformations acquire such a form when the RGS
are of the Lie--B\"acklund group type,\footnote{The set of group
variables of the generalized Lie group of transformations (the
Lie--B\"acklund group) includes derivatives of the desired
functions w.r.t.\ independent variables.}\ and the FS condition
claims that the particular solution of a boundary-value problem is
invariant w.r.t.\ these transformations.

In Sec.~5, we analyze the FS conditions for systems with small
parameters. The FS conditions in such systems permit finding
approximate RGSs and using them to solve problems with arbitrary
boundary conditions. We present examples of approximate RGSs for
several problems in nonlinear physics.

In Sec.~6, we describe the perspectives of the FS and its use to construct
RGSs in a broad class of mathematical physics problems.

\section{The QFT RG and the FS condition}

\subsection{Simple Bogoliubov RG transformations}

To illustrate the FS notion, we consider the simplest RG transformation,
which is a simultaneous one-parameter transformation of one independent
variable in a problem (the coordinate $\ell$, for example) and a
characteristic~$g$ of a solution,
\begin{equation}  \label{rgtr-a}
 T (\lambda) : \left\{\ell \to \ell' = \ell - \lambda~,~
 g \to g'= G(\lambda, g)~\right\}~, \quad G(0, g)=g.
\end{equation}
The function $G(\ell,g)$ satisfies the functional relation
\begin{equation}   \label{feq-a}
 G(\ell + \lambda , g) = G(\ell, G(\lambda, g)),
\end{equation}
which corresponds to the group composition law for the
transformation operators,
 $T(\lambda_1)\cdot T(\lambda_2) = T(\lambda_1 + \lambda_2)$.

Transformation operator~(\ref{rgtr-a}) can sometimes be
conveniently represented explicitly by writing a transformation of
a function $F(\ell, g)$ in the form
\begin{equation}  \label{globf-a}
 T(\lambda)\, F(\ell, g) = e^{- \lambda R}\, F(\ell, g)  \,
\end{equation}
using the infinitesimal RGS operator (or the RG operator)
\begin{equation}  \label{rg-op}
R=\partial_{\ell}-\beta(g)\partial_g \, , \quad \beta(g)={\frac{\partial
G(\lambda; g)} {\partial \lambda}}      {\bigg\vert_{\lambda = 0}} \,,
\end{equation}
whose coordinate $\beta(g)$ is the derivative of the function~$G$
at $\lambda=0$. The infinitesimal RG transformations can be
conveniently represented via the operator~$R$; the group nature of
the operator $T(\lambda)=e^{-\lambda R}$ is obvious from ``finite
shift" representation~(\ref{globf-a}).

The condition
\begin{equation}  \label{diff-a}
  R  I(\ell, g) \equiv \partial_{\ell}I -\beta(g)\partial_g I = 0 \,,
\end{equation}
determines an {\sl invariant} of the RG transformation, which is
the function $I(\ell,g) = \tilde I(G(\ell,g))$ $ \equiv
 I\left(0, G(\ell, g)\right)$ of one argument in this case because
the transformation function $G(\ell,g)$ is an invariant itself by
virtue of Eq.~(\ref{feq-a}). The condition
\begin{equation} \label{covar-a}
 R\,C(\ell, g) =\varphi(g) C(\ell, g)
\end{equation}
determines a {\sl covariant}, i.e., a quantity that transforms
according to a representation of the FS group or the RG.
Covariants are important for QFT applications of the RG.

The representation of a RG transformation via an infinitesimal
operator~$R$ is equivalent to finite
transformation~(\ref{globf-a}). Functional equation (\ref{feq-a})
follows from the characteristic equation for the operator~$R$, and
for a given function $\beta$, an explicit expression for~$G$ can
be constructed by solving the corresponding Lie equations for
operator (\ref{rg-op}),
\begin{equation}  \label{Lie1}
-d \ell' =\frac{d g'}{\beta(g')}= d \lambda
\end{equation}
with the boundary conditions $\quad\ell'\vert_{\lambda =0}=\ell$
and $g'\vert_{\lambda=0}=g$.

It is important in what follows that the invariance of the
function~$I$ w.r.t.\ the RG transformations written as
Eq.~(\ref{diff-a}) is equivalent to the vanishing condition for
the coordinate~$\varkappa$ of operator~(\ref{rg-op}), which is
written in the canonical form as
\begin{equation}   \label{canon1}
 \bar R=\mbox{\ae}\partial_{I}\,, \quad
 \mbox{\ae} \equiv  {I}_\ell - \beta(g) {I}_g = 0 \,.
\end{equation}
This condition must be considered on a particular solution
$I(\ell,g)$ of the boundary-value problem.

The ``exponentiated" variables
 $$
 x=e^{\ell},\qquad  a=e^{\lambda},\qquad\bar{g}(a,g)=G(\lambda,g),
 $$
are natural in QFT. In these variables, the group transformation
$T(\lambda)= T_a$ becomes
\begin{equation}  \label{rgtr-m}
 T_a : \left\{ x' = x/a~,~ g'=\bar{g}(a, g)~\right\}~,\quad
       \bar{g}(1, g)=g\,,
\end{equation}
and
 \begin{equation}   \label{feq-m}
 \bar{g}(x,g)= \bar{g}\left(x/a, \bar{g}(a, g)\right), \quad
 T_a\cdot T_b = T_{ab}.
\end{equation}
Functional equation~(\ref{feq-m}) and
transformation~(\ref{rgtr-m}) appear, for example, in a QFT with
one coupling constant in the massless limit, where the
dimensionless constant $x=Q^{2}/\mu^{2}$ is the ratio of the
squared transferred four-momentum~$Q$ to the squared ``normalized"
momentum~$\mu$, $g$ is the coupling constant, and the
invariant~$\bar{g}$ is the invariant (or effective) coupling
function. More complicated QFT RG transformations are
generalizations of transformation~(\ref{rgtr-m}) obtained by
``multiplying" solution characteristics,
$g\to\{g\}=(g_1,g_2,\dots,g_k)$, and by introducing additional
parameters in the functions $\bar{g}_i(x,\{g\})$ (see Sec.~49 in
\cite{Book}).

In the QFT model with two coupling constants $g_1=g$ and $g_2=h$
and one particle mass~$m$, for example, the transformation $T_a$
becomes
\begin{equation} \label{rgtr-m2}
T_a :\left\{ x'= x/a~,\ ~ y'= y/a~,~\ g'=\bar{g}(a, y; g, h)~, \
~h'= \bar{h}(a, y; g, h)~\right\}~\, ,
\end{equation}
 $$ \bar{g}(1, y; g, h)=g\,, \quad \bar{h}(1, y; g, h)=h\,,\quad
 y=Q^2/\mu^2 \, .
 $$
In the case of several coupling constants, the characteristic equations
that are connected with the infinitesimal operator
\begin{equation}  \label{rg-op2}
R = x\partial_x + y\partial_y-\beta_1(y; g, h)\partial_g
-\beta_2(y; g, h)
\partial_h \,,
\end{equation}
$$
 \beta_1 (y; g, h)={\frac{\partial \bar{g}(\xi, y; g, h)}{\partial \xi}}
         {\bigg\vert_{\xi = 1}} \,, \quad
 \beta_2 (y; g, h)={\frac{\partial \bar{h}(\xi, y; g, h)}{\partial  \xi}}
         {\bigg\vert_{\xi = 1}} \,,
$$
become the system of first-order PDEs
\begin{equation} \label{deq-sys2}
 x\, \frac{\partial \bar{g}(x, y; g, h)}{\partial
 x}=\beta_1(\frac{y}{x}; \bar{g}, \bar{h}) \, , \quad
 x\, \frac{\partial \bar{h}(x, y; g, h)}{\partial x}= \beta_2
 (\frac{y}{x}; \bar{g}, \bar{h}) \, ,
\end{equation}
however, all these one-parameter transformations are based on the
transformation of a single independent argument~$x$.

\subsection{Self-similarity and the FS}
In a particular case where the function~$\bar{g}$ is linear in its
second argument, $G\sim g$, solutions of Eq.~(\ref{feq-m}) have a
powerlike dependence on the argument~$x$, i.e.,
$\bar{g}(x,g)=gx^{k}$, where~$k$ is a number, and
transformations~(\ref{rgtr-m}) become transformations of powerlike
self-similarity (the so-called automodel transformations)
 $$ P_a : \left\{ x'=x/a \,, \quad g'= g a^{k} \right\} \,, $$
which are commonly used in problems in gas and liquid dynamics.
From this standpoint, transformations~(\ref{rgtr-m})~and
(\ref{feq-m}) for arbitrary $G,\bar{g}$ are functional
generalizations $gx^k\to\bar{g}(x,g)$ of customary self-similarity
transformations and can therefore be called the {\sl functional
self-similarity} transformations~ \cite{dv82}; this term is a
synonym for the RG transformations.

Therefore, in the RG-transformation framework, FS reflects the
group nature of functional relations. The universality of the FS
formulation in the QFT and classical physics models is due to the
common scaling transformation and common functional transformation
of the solution characteristic $g_{\mu}= \bar{g}(\mu,g)$. Various
realizations of the RG differ only in the form of the function
$\beta(g)$, which is customarily calculated using an approximate
solution obtained, as a rule, within the perturbation theory (PT).
Therefore, the main role of the FS until recently was to establish
a~posteriori that a system under consideration admits functional
transformations with a group structure. Group transformations
themselves followed from additional considerations concerning
system solutions~\cite{BBSh-JETF57}.

\subsection{Constructing the RGS}
The situation changed substantially after passing to investigating
boundary-value problems in mathematical physics. The FS property of a desired
solution here results in functional relations (which follow from the group
nature of a solution) or in symmetries of the RG (=FS) type in the
infinitesimal form. In contemporary mathematical physics models, RGSs can be
regularly found by using the scheme in~\cite{KPSh-JMP98}, which naturally
incorporates the FS and its formulation as the invariance condition for a
particular solution of a boundary-value problem.

To demonstrate the role of FS in constructing RGSs, we recall the
four main steps in this construction. The first step is to
construct a special RG manifold (differential,
integral-differential, etc.), which {\sl differs}, generally
speaking, from the manifold determined by the initial system of
equations describing the physical system under investigation. The
second step is to find the maximum extended transformation group
$\cal{G}$ admitted by the RG manifold. The third step is to
restrict the obtained group $\cal{G}$ on a solution (exact or
approximate) of a boundary-value problem. The transformation group
that appears, which is just called the RG, is a set of
infinitesimal operators~$R_i$, each of which contains a solution
of the boundary-value problem in its invariant manifold. The
fourth and final step is to use the infinitesimal RG operators
$R_i$ to construct finite transformations of the group and to
obtain an analytic expression for the solution of the
boundary-value problem.

The first and principal step in this scheme for RGS construction
can be realized in different ways depending on the mathematical
model structure and on the type of boundary
conditions~\cite{KPSh-JMP98}. The desired RG manifold can be
obtained by adding parameters that enter the initial equations and
boundary data to extend the list of group variables or appending
to this list derivatives of the given dependent variables or
nonlocal variables. We can also increase the number of initial
equations by writing boundary data either in the embedding
equation form or as additional differential constraints.
Sometimes, the admitted group can be extended by dropping small
parameters to simplify the initial equations.

The maximum group $\cal{G}$, which is calculated for the RG
manifold using the modern group analysis algorithms~\cite{Oves} in
the second step, is not yet the RG, because it is not connected
with a particular solution.

The third step constructs the RG itself; namely, we restrict the
obtained group $\cal{G}$ on a particular exact or approximate
solution of the selected boundary-value problem. Mathematically,
the restricting procedure implies ``combining" the canonical
coordinates of operators of the group admitted by the RG manifold.
The vanishing condition for the sum of these coordinates on the
solution of the boundary-value problem (the condition of
invariance w.r.t.\ the RGS operator, which is analogous to
condition~(\ref{canon1}))) results in a system of algebraic
identities, which relate coordinates of different operators and
therefore generate the desired RGSs. From the physical standpoint,
the verification of this condition in each actual case is the
realization of the FS principle, which therefore establishes the
invariance of the particular solution w.r.t.\ the RG
transformations. On the other hand, this step substantially relies
on the PT solution of the problem under investigation
simultaneously realizing the general principle of the RG method
consisting in constructing an improved (in comparison with the PT)
solution. In this scheme, the PT parameter can be any physically
appropriate parameter or a set of such parameters.

Verification of the FS condition in the RGS construction scheme
above differs from the procedure for verifying the invariance of
the analyzed solutions w.r.t.\ an iterated sequence of scaling
transformations, which are used to analyze the behavior of a
physical system in the theory of critical phenomena \cite{ken72}
and are also called RG transformations although they do not
constitute a group in the general case (in contrast to the QFT
models, where the group property is especially
proved~\cite{Book,BBSh-JETF57}). This scheme (we call it the
Wilson RG) is now widely used in mathematical physics to analyze
the asymptotic behavior of DE solutions~\cite{gold96,Bric94} and
to construct the envelope of the solution family~\cite{Kunih}.
When applying the Wilson RG to mathematical physics problems, the
algorithm for improving the PT solutions that contain
singularities consists of introducing additional parameters in the
solutions, using these parameters to remove singularities, and
demanding the solutions to be independent of the way these
parameters are introduced \cite{gold96}. Whether such a
construction is consistent with the transformation group of the
desired boundary-value problem solution remains an open question
although this algorithm works successfully in some particular
cases.

The RGS in mathematical physics is defined by a set of RGS
generators. Therefore, as in the classical group analysis of DEs,
it suffices to consider only infinitesimal transformations that
are characterized by an infinitesimal operator algebra. We note
that this algebra often consists of more than one operator in
contrast to the QFT-type models where we typically have only one
RG operator. Both the dimension and the construction of the RG
operator algebra depend on the mathematical model and on the type
of boundary conditions.

The RGS group, which is restricted on a solution of a
boundary-value problem, can be not only a point Lie group but also
a Lie--B\"acklund group, an approximate transformation group, a
nonlocal symmetry group, a non-Lie symmetry group,
etc.~\cite{KPSh-JMP98}. We present the FS conditions and various
forms of the FS transformations pertaining to concrete problems in
which the RGS is used.

\section{The FS analysis of systems that are close to quantum field systems}
\label{sec3}

We begin with the FS transformations for systems described by ODEs
and, for simplicity, consider a boundary-value problem for the
function $u(t)$ satisfying a first-order
ODE~\cite{KPSh-PD447-95,KKP-RG91,KPSh-JMP98} with the parameters
$b^k$ ($k=1,2,\dots$) explicitly entering the equation:
\begin{equation}  \label{odu1}
 u_t=f(t,u,b^k)\,,
\end{equation}
\begin{equation}  \label{odu2}
u(\tau)=x\,.
\end{equation}
This example is not only methodologically important; it illustrates the FS
application to QFT where such equations appear in the RG method when
analyzing a system of DEs for invariant coupling functions.

Constructing the RGS for boundary-value problem~(\ref{odu1}), (\ref{odu2})
using the algorithm in~\cite{KPSh-JMP98} consists in adding to~(\ref{odu1})
the embedding equation, which has the form of a linear first-order PDE,
\begin{equation}  \label{pogr1}
 u_{\tau}+f(\tau,x,b^k)u_x=0 \,.
\end{equation}
The system of equations~(\ref{odu1}) and~(\ref{pogr1}) determines
the desired RG manifold in the space of all group parameters
$\{u,t,\tau,x,b^k\}$, which are not only dependent and independent
variables and the parameters of initial equation~(\ref{odu1}) but
also the parameters~$\tau$ and~$x$ entering the boundary data.

The symmetry group $\cal{G}$ admitted by
manifold~(\ref{odu1}),~(\ref{pogr1}) depends on the form of the
function~$f$. In the practically important case (the ultraviolet
limit of QFT models) where this function does not depend
explicitly on time~$t$ and only three parameters $b^1\equiv a$,
$b^2\equiv b$, and $b^3\equiv c$ enter the initial equation, i.e.,
where $f=f(u,a,b,c)$, the admitted group $\cal{G}$ is determined
by the seven-term operator
\begin{equation}   \label{odugr}
 X= \sum\limits_{i=1}^{7}\alpha_i X_i\,,
\end{equation}
The first two functions $\alpha_1$ and $\alpha_2$ in~(\ref{odugr})
are arbitrary functions of all seven group variables
$\{t,\tau,x,a,b,c,u\}$, and the remaining functions depend
arbitrarily on the parameters~$a$, $b$, and $c$ and on the
combinations $\tilde t=t-\langle 1/f(u)\rangle$ and
$\tilde\tau=\tau-\langle 1/f(x)\rangle$. The angle brackets denote
integrals w.r.t.\ the respective variable~$u$ or~$x$. Explicit
expressions for three of the seven operators
entering~(\ref{odugr}) are
 $$ \begin{array}{c}
 \displaystyle{ X_1= \partial_t + f(u) \partial_u \,, \
                X_3=  f(u) \partial_u \,,}
 \\ \mbox{} \\
 \displaystyle{
  X_5=  f(x)<f_a(x)/f^2(x)>\partial_x +
        f(u)<f_a(u)/f^2(u)>\partial_u +
                           \partial_a \,.}\\
 \end{array}         $$
The remaining operators are obtained by substitutions: $X_2$ and $X_4$ are
obtained from $X_1$ and $X_3$ with the respective substitutions $t\to\tau$
and $u\to x$, and $X_6$ and $X_7$ from $X_5$ with the respective derivative
substitutions $\partial_a\to\partial_b$ and $\partial_a\to\partial_c$.

The operation of restricting the group $\cal{G}$ is the
verification of the imposed FS condition, which is analogous to
the equality $\varkappa=0$ in (\ref{canon1}), i.e., the solution
must be invariant w.r.t.\ the RG transformations, or, in other
words, the coordinate of the canonical operator~$X$ must vanish on
the solution of the initial problem $u= U(t,x,\tau,a,b,c)$. If a
restriction of the group $\cal{G}$ admitted by
manifold~(\ref{odu1}),~(\ref{pogr1}) on an (approximate) solution
of problem (\ref{odu1}),~(\ref{odu2}) is fulfilled, the
``restricted symmetries," which we call RGSs,\footnote{Some $R$
operators of RGSs thus defined may result from symmetries of the
equations.}\ appear.

Which perturbation series becomes the function~$U$ depends on the actual
problem setting. For instance, for a polynomial function~$f$,
\begin{equation} \label{tripod}
 f=au^2+bu^3+cu^4
\end{equation}
we can choose the PT over the variables $(t-\tau)$ or over the
parameter~$a$, $b$, or~$c$. Examples of restricting the group on a
solution with a PT in~$a$ with $b=c=0$ was considered
in~\cite{KKP-RG91,KPSh-JMP98}, and with a PT in~$b$ with $a=1$ and
$c=0$ in~\cite{KPSh-JMP98}. We present two possible RGS operators
appearing after such restrictions~\cite{KKP-RG91,KPSh-JMP98},
 \begin{equation}   \label{rg1} \begin{array}{c}
 \displaystyle{
 R_1= x^2 \tau \partial_x + \partial_a + u^2 t \partial_u \,, \ b=c=0\,,}
 \\ \mbox{} \\
 \displaystyle{   R_2 = \left( x^2(1+bx)\tau+x \right) \partial_x
     + \left( u^2 (1 + b u)t + u \right) \partial_u
     - b \partial_b  \,, \ a=1\,, \ c=0\,.} 
 \end{array}
\end{equation}
We can now obtain the solution of the Cauchy problem if we write
the condition for its invariance w.r.t.\ the RGS operator, namely,
the FS condition, which is a linear first-order PDE. For instance,
for the operator $R_1$, we have
 \begin{equation}  \label{fss1}
 t u^2 - x^2 \tau u_x - u_a = 0\,.
\end{equation}
Solving the characteristic equations (the Lie equations) for this
equation, we obtain the desired solution of the boundary-value
problem for $b=c=0$. The FS condition of type~(\ref{fss1}) is used
twice: first, when constructing the RGS operators, we substitute
an approximate solution~$U$ for~$u$ in (\ref{fss1}); second, when
finding the solution of the boundary-value problem, we use these
RGS operators.

In both examples, constructing the PT is easy and results either
in the powerlike dependence or in a combination of powerlike,
reciprocal, and logarithmic dependencies on the initial data. In
this algorithm, constructing the PT in powers of~$b$, which gives
the second operator $R_2$ in~(\ref{rg1}), starts from the
unperturbed state, which is chosen to be the solution of the
boundary-value problem at $a=1$ obtained by applying the first
operator $R_1$, i.e., in the approximation of strong nonlinearity
(w.r.t.\ the parameter~$a$). Therefore, the improvement of the PT
using the operator $R_2$ is reduced to the consequent improvement
of the PT for the boundary-value problem with the function
$f=au^2+bu^3$ first in the parameter~$a$ and then in the
parameter~$b$ using the respective one-parameter groups~$R_1$
and~$R_2$.

In many cases, such a procedure of improving the PT consecutively
over several parameters of a model fails. One possible reason is
the absence of a ``nonlinear" unperturbed solution. For the
function~$f$ discussed here, which is chosen as a polynomial
in~$u$, constructing the PT in~$c$ with fixed~$a$ and~$b$ is not
so simple as constructing the PT at $c=0$, because in the case
with fixed $a$ and $b$, the unperturbed state is determined in
terms of the Lambert function~\cite{Maple3-5}, which admits no
simple analytic representation (see also~\cite{dvsol-tmp99}).

Therefore, constructing an RGS that generates group transformations w.r.t.\
several parameters simultaneously and improves a PT that admits an easily
found unperturbed solution is important. In fact, we want to construct a
two-parameter RG. This problem will be discussed in detail elsewhere; we
consider only one example of such an RG here:
\begin{equation} \begin{array}{c}
 \displaystyle{
  R_3=  - <f_b(u)/f^2(u)>\partial_t
        - <f_b(x)/f^2(x)>\partial_\tau + \partial_b \,,}
 \\ \mbox{} \\
 \displaystyle{    R_4=  - <f_c(u)/f^2(u)>\partial_t
        - <f_c(x)/f^2(x)>\partial_\tau + \partial_c \,.}\\
 \end{array}
\end{equation}
Verifying the FS conditions, i.e., restricting the group $\cal{G}$
on the PT solution over the two parameters $b\to 0$ and $c\to 0$
simultaneously is not difficult technically, and using the finite
transformations generated by the operators $R_3$ and $R_4$, which
generate a two-dimensional algebra, we obtain the desired solution
of problem~(\ref{odu1}),~(\ref{odu2}). Therefore, using the
two-parameter RG, we can avoid mathematical problems that arise
when using one-parameter RGs.

We conclude this section with a few remarks. We have demonstrated
the use of the FS condition to construct RGSs that improve the PT
over the parameters $a$, $b$, and~$c$ entering the equation,
although the corresponding operators by no means exhaust an
infinite set of RGSs, which is parameterized by a continuum set of
the RG operator
coordinates~\cite{KPSh-PD447-95,KKP-RG91,KPSh-JMP98}. Quite
analogously, RGS operators that improve a PT w.r.t.\ an
appropriate combination of dynamic variables and initial data
(e.g., w.r.t.\ the difference $t-\tau$) can be found. Examples of
such RGSs are presented in the following sections. We only mention
here that the FS condition in the form of equality~(\ref{canon1})
is obviously analogous to FS condition (\ref{fss1}) pertaining to
boundary-value problem~(\ref{odu1}),~(\ref{odu2}). The RGS
examples and the corresponding FS condition considered in this
section follow from a single ODE. Similar constructions are valid
for systems of ODEs depending on several parameters. Examples of
embedding equations for such systems, which were used to find
RGSs, can be found in~\cite{KKP-RG91}.

\section{The FS conditions for PDEs}

In this section, we consider systems that are described by
mathematical models based on PDEs or systems of PDEs. In contrast
to the previous section, where the appearance of first-order
partial derivatives in FS conditions is due to a transformation of
parameters in initial equations and due to embedding equations,
which imply taking the boundary conditions into account, we here
demonstrate boundary-value problems for which the FS conditions
are systems of DEs (constraints) that contain higher-order partial
derivatives w.r.t.\ the independent variables. The differential
formulation of the FS conditions then permits constructing the
desired solution of the problem; finding finite FS transformations
results in formal power series (see \cite{Sprav}, Vol.~3,
Chap.~1). In addition, we give an example of an FS condition
(simpler than in Sec.~3) in the ODE form whose solution also
reconstructs the solution of the boundary-value problem from the
given PT.

\subsection{Boundary-value problem for the Burgers equation}
An explicit example, which illustrates the variety of the FS condition
formulations for systems based on PDEs, is provided by the boundary-value
problem for the modified Burgers equation,
\begin{equation} \label{bureq}
 u_t-au_x^2-{\nu}{u}_{xx}=0\,,
\end{equation}
\begin{equation}  \label{burcond}
  u(0,x)=f(x) \,,
\end{equation}
with the nonlinearity parameter~$a$ and dissipation
parameter~$\nu$. The continuous point symmetry group admitted by
manifold~(\ref{bureq})  is defined by nine operators. Six of them
are symmetries of the equation and have been discussed in the
literature (see, e.g.,~\cite{Sprav}, Vol.~1, Chap.~1, p.~183).
They correspond to the projective transformation, the dilation
transformation in the plane $(t,x)$, translations along the
axes~$t$, $x$, and~$u$, and the Galilean transformations. The
seventh operator is the operator of an infinite Abelian ideal
$X_{\infty}=\alpha\exp(-{au}/{\nu})\partial_u$ of the group; the
coordinate of this ideal is parameterized by the function
$\alpha(t,x,a,\nu)$ of four group variables restricted by the
linear parabolic equation $$ \alpha_t-\nu\alpha_{xx}=0, $$ which
coincides with the linear part $(a=0)$ of initial
equation~(\ref{bureq}). Eventually, when we interpret the equation
parameters as independent variables, two more operators appear and
involve these parameters in the group transformations; these
operators correspond to scaling transformations of the respective
variables $a$ and~$\nu$~\cite{KP-LGA94}.

Restricting the group admitted by manifold~(\ref{bureq}) on the
solution $u=U(t,x,a,\nu)$ of the Cauchy problem, i.e., verifying
the FS condition, results in an algebraic relation, which
expresses the coordinate of the infinite-dimensional subgroup
generator (the function~$\alpha$) through the coordinates of the
remaining eight operators at any time~$t$, including $t=0$ when
this solution $U(0,x,a,\nu)=f(x)$ is known from boundary condition
(\ref{burcond}). Using the standard representation for the
solution of a linear equation on the function~$\alpha$ with the
initial value $\alpha (0, x,a,\nu)$ obtained from the FS condition
and substituting this representation in the formula that
determines the general element of the Lie algebra, we obtain the
desired RGS operators. Thus, we obtain the RGSs for boundary-value
problem (\ref{bureq}),~(\ref{burcond}) by combining the symmetries
of the eight-dimensional algebra generated by the above operators
and symmetries of the infinite-dimensional subalgebra generated by
the operator $X_{\infty}$. Each of the eight RG operators obtained
(and their linear combinations whose coefficients are arbitrary
functions of~$a$ and~$\nu$) contains a solution of the Cauchy
problem $u=U(t,x,a,\nu)$ in the invariant manifold and permits
finding group transformations of both the variables $\{t,x,a,\nu
\}$ and various functionals (local and nonlocal) of the solution.

We present two such RGS operators, which improve the corresponding
PTs over the parameter~$a$ (operator $R_5$) and the independent
variable~$t$ (operator $R_6$),
\begin{equation}
R_5= \partial_a + \frac{1}{a}\left( -u + \exp \left( -\frac{au}{\nu} \right)
   <f(x)>  \right) \partial_u \,, \label{burrg1} \end{equation}
\begin{equation}   \label{burrg2}
 R_6 =\partial_t +  \exp \left( -\frac{au}{\nu} \right)
  <af_x^2 + \nu f_{xx}>  \partial_u \,.
\end{equation}
The double angle brackets here are integral convolutions of the
corresponding functions with the fundamental solution $G(t,x,\nu)$
of the linear equation for the function~$\alpha$ multiplied by the
exponent of the function~$f$ from boundary
condition~(\ref{burcond}):
  $$
 <F(x)>\equiv
 \frac{1}{\sqrt{4\pi \nu t}}\int\limits_{-\infty}^{\infty}\,dy\,
 F(y)  \exp \left(-\frac{(x-y)^2}{4 \nu t} + \frac{a f(y)}{\nu}\right) \,.
 $$
The invariance conditions for a solution of a boundary-value
problem (the FS conditions) corresponding to RG
operators~(\ref{burrg1}) and~(\ref{burrg2}) are the two
first-order ODEs
\begin{equation}
 -u_a - \frac{u}{a} +
 \frac{1}{a}\exp \left( -\frac{au}{\nu} \right)  <f(x)> =0 \,,
 \label{burfss1}
\end{equation}
\begin{equation}
 -u_{t} +  \exp \left( -\frac{au}{\nu} \right)
  <af_x^2 + \nu f_{xx}> =0 \,.
 \label{burfss2}
\end{equation}
Approximate solutions of Eq.~(\ref{bureq}) can be extended in
parameters~$a$ or $t$ by solving either Eq.~(\ref{burfss1})
or~(\ref{burfss2}) with the proper initial conditions. This
eventually leads to the common exact solution $u=(\nu/a)\ln
\langle\langle 1\rangle\rangle$, which is valid for all values of
the parameters~$a$ and~$t$~\cite{KP-LGA94}.

In the example above, FS condition~(\ref{burfss1}) for the
boundary-value problem for a PDE becomes a first-order ODE.
Formally, this equation is simpler than the FS conditions
formulated for the ODE boundary-value problems in Sec.~3 and
differs from the corresponding FS conditions for QFT models. On
the other hand, formulating boundary conditions for a first-order
ODE in the embedding equations language, we notice that
condition~(\ref{burfss1}) can be treated as an embedding equation
for boundary-value problem~(\ref{bureq}),~(\ref{burcond})  with
the embedding parameter~$a$. In turn, the invariant embedding
method can also be applied to boundary-value
problem~(\ref{burfss1})  with the variables $\{a,u\}$ and the
initial condition $u=u_0(t,x,\nu)$ at $a=0$. Then, because the
initial value $u_0$ depends on the ``parameters" $\{t,x,\nu\}$
entering ODE (\ref{burfss1}), the corresponding embedding equation
becomes integral-differential~\cite{KPSh-PD447-95}. The set of the
RGS operators for boundary-value
problem~(\ref{bureq}),~(\ref{burcond})  includes not only the
above FS conditions in the form of first-order ODEs but also
operators that result in FS conditions both in the form of
first-order PDEs, which are analogous to the operators previously
considered, and in the form of mere algebraic
relations~\cite{KP-LGA94}. Therefore, various FS can relate to the
solution of the same boundary-value problem.

\subsection{The boundary-value problem for nonlinear optic equations
\label{4.2}}

It is sometimes impossible to make a boundary-value problem PT and
the FS conditions consistent if we confine ourselves to only point
symmetries. Below, we consider an example where the FS condition
is a second-order PDE. We now consider the following
boundary-value problem for the system of equations for the
nonlinear optics of collimated wave beams:
\begin{equation}  \label{nlopeq1}
v_t + v v_x - \alpha n_x=0\,, \quad n_t + vn_x + nv_x = 0 \,,
\end{equation}
\begin{equation}
 v(0,x) = 0 \,, \quad n(0,x) = N(x)\,.
\label{nlopcond1}
\end{equation}
Here, the dimensionless coordinates~$t$ and~$x$ describe the
spatial evolutions of the derivative~$v$ of the beam eikonal and
of the dimensionless intensity~$n$ in the direction into the bulk
and in the transverse direction, and $\alpha$ is the parameter of
the nonlinear refraction. The hodograph transformation
reduces~(\ref{nlopeq1})  to the system of linear equations in the
variables $\tau=nt$ and $\chi=x-vt$
\begin{equation} \label{nlopeq2}
 \tau_w - n \chi_n = 0\,, \quad \chi_w + \alpha \tau_n = 0 \,,
\end{equation}
\begin{equation}
 \tau(0,n)=0 \,, \quad \chi(0,n)=H(n)\,.
\label{nlopcond2}
\end{equation}
where we use the notation $w=v/\alpha$. Formal construction of the
RGS and analysis of the FS conditions for system~(\ref{nlopeq2})
could be as for the Burgers equation. As previously, a
finite-dimensional subalgebra of point symmetry operators and its
infinite-dimensional ideal arise. However, the procedure of
restricting this ideal in order to obtain the
RGS~\cite{KP-MCM97,K-JNMP96} implies solving the system of linear
PDEs, which coincides with initial system (\ref{nlopeq2}); this is
an intrinsic feature of linear equations.

Therefore, we construct the RGS using not the point symmetries but
the Lie--B\"acklund symmetries (see \cite{Sprav}, Vol.~3,
Chap.~1), for which the terms ``higher" and ``generalized"
symmetries are also used. The desired RGS operators for
boundary-value problem~(\ref{nlopeq2}),~(\ref{nlopcond2}) can then
be conveniently written in the canonical form
\begin{equation}   \label{oprgnl1}
R=f \partial_{\tau} + g \partial_{\chi} \,,
 \end{equation}
where the coordinates~$f$ and~$g$ are functions in the extended
space of group variables, which, in addition to the set
$(\tau,\chi,w,n,\alpha)$, includes derivatives of~$\tau$
and~$\chi$ of arbitrary finite order in~$n$ (and, perhaps,
in~$\alpha$). The problem of constructing the Lie--B\"acklund
symmetries for boundary-value
problem~(\ref{nlopeq2}),~(\ref{nlopcond2})  was discussed in
detail in~\cite{KP-MCM97}. It was shown that these symmetries are
generated by operators of form~(\ref{oprgnl1})  whose coordinates
are linear combinations of $\tau$,~$\chi$, and their first- and
second-order derivatives depending on $w$ and $n$. We write the
expressions for the coordinates of only three operators:
 \begin{equation} \label{oprgnl2}
 \begin{array}{l}
 \displaystyle{ f_1=-\tau /2 + n \tau_n + (1/2)n w\chi_n,
 \quad g_1=-(\alpha w/2)\tau_n + n\chi_n;  }\\ \mbox{} \\
 \displaystyle{
 f_2=n\tau_n, \qquad g_2=\chi_n + n\chi_{nn} ;}\\ \mbox{} \\
 \displaystyle{
 f_3=(1/4)\tau-n\tau_{n}-(5/4)wn\chi_n +\left(-n+(\alpha/4)w^2
    \right)n\tau_{nn} - wn^{2}\chi_{nn}, } \\ \mbox{} \\
 \displaystyle{
 g_3=(3/4)v\tau_n -\left(2n -(\alpha/4)w^2\right)\chi_n +\alpha
      wn\tau_{nn} +
 \left(-n + (\alpha/4)w^2\right)n\chi_{nn} \,.} \end{array}
 \end{equation}
The first operator with the coordinates~$f_1$ and~$g_1$, which are linear in
the first derivatives, is equivalent to the point symmetry operator; the
other two operators are the Lie--B\"acklund symmetry operators of the second
order. Restricting the Lie--B\"acklund group found for the RG manifold
consists in verifying the FS condition
\begin{equation}   \label{fssnl1}
f=0\,, \quad g=0 \,,
 \end{equation}
which must be satisfied on the solution of the boundary-value
problem and be consistent with boundary
conditions~(\ref{nlopeq2}). The form of the coordinates $f$
and~$g$ implies that these conditions, written in terms of the
boundary function $H(n)$, must satisfy linear PDEs with variable
coefficients. In particular, for the case $N(x)=\cosh^{-2}(x)$,
the FS condition is fulfilled for the combinations
$f=f_1+2(f_2+f_3)$ and $g=g_1+2(g_2+g_3)$, and the desired
Lie--B\"acklund RGS operator is
\begin{equation}  \label{rgs1-sol}
\begin{array}{l} \displaystyle{R_7 = \left(2n(1-n)\tau_{nn}- n\tau_n -
2nw(\chi_n +n\chi_{nn})+\frac{\alpha}{2}\,nw^2\tau_n\right)\partial_{\tau}}
\\ \mbox{}\\
\displaystyle{ + \left( 2n(1-n)\chi_{nn} +(2-3n)\chi_n + \alpha w \left(
2n\tau_{nn} + \tau_n + \frac{ w}{2}\left( n \chi_{nn} + \chi_n \right)
    \right) \right) \partial_{\chi} \,. } \end{array}
 \end{equation}
Condition~(\ref{fssnl1})  for the operator $R_7$ results in a
second-order PDE, which distinguishes it qualitatively from the
first-order equations arising in the FS conditions considered
previously. Namely, the finite FS transformations for the latter
can be found in a closed form as the solutions of the
corresponding Lie equations expressing the transformations of
dependent and independent variables and parameters. The solutions
of the Lie equations for the RGS with Lie--B\"acklund
operator~(\ref{rgs1-sol}) result in finite transformations written
as formal power series. However, this does not mean that FS
conditions have no practical importance in this case; on the
contrary, the differential formulation of the FS conditions
permits considering relations~(\ref{fssnl1})  as additional
differential constraints, which must be taken into account when
finding solutions of the boundary-value problem. The solution of
the initial equations with the differential constraints taken into
account leads to the solution of the boundary-value
problem~\cite{KP-MCM97,K-JNMP96,KSh-JNOPM97} that is invariant
w.r.t.\ RGS group~(\ref{rgs1-sol}).

To conclude this section, we note that the FS conditions written
as vanishing conditions for coordinates of a Lie--B\"acklund
canonical RG operator, being considered as a system of
differential constraints, play the role of embedding equations.
Imposing the FS conditions, we come to the problem of finding the
solution that is invariant w.r.t.\ the found RGS operator. A more
informative example is provided by using the FS conditions to find
point RGS operators as the group admitted by the initial equations
and FS relations~\cite{KPSh-PD447-95}. There, RGS operators
containing higher-order derivatives w.r.t.\ parameters entering
the equations can be discussed in principle.

\section{The FS and RGS for systems with small parameters \label{sec5}}

We now consider the FS property for systems described by DEs with
small parameters. If a system contains a small parameter~$\alpha$,
we can start the RGS construction by considering the simplified
($\alpha=0$) model, which admits a wider symmetry group in
comparison with the case $\alpha\neq0$. When we take the
contributions from small~$\alpha$ into account, this symmetry is
{\sl inherited} by the initial system of equations, which results
in the appearance of additional terms, corrections in powers
of~$\alpha$, in the operator coordinates. Restricting the obtained
symmetry group to an exact or approximate solution of the
boundary-value problem, we obtain the desired RGS, which can be
also represented as operators whose coordinates are infinite
series in powers of small parameters. For boundary data of a
special form, these series terminate, which results in an exact
RGS without restrictions on values of the relevant parameters. For
boundary data of a general type, the procedure of truncating the
infinite series to a finite number of terms leads to an
approximate RGS at small parameter values.

\subsection{FS conditions in nonlinear plasma theory}

We now consider the boundary-value problem for the equations of
the nonlinear interaction of laser radiation with a
plasma~\cite{KP-TMF89}. Such an interaction for a $p$-polarized
electromagnetic wave of frequency~$\omega$, for which only the $z$
component of the magnetic field is nonzero and which propagates
from the vacuum toward an inhomogeneous plasma, is described by a
system (we do not present it here) of nonlinear nonstationary
equations for six scalar functions: two components of the electron
velocity~$v_x$ and $v_y$, the electron density~$n$, two electric
field components~$E_x$ and~$E_y$, and the $z$ component $B_z$ of
the magnetic induction; these functions depend on the time $t$ and
the two coordinates~$x$ and~$y$.

The nonlinearity of these equations is essential in a small space
domain near the plasma resonance (at $\omega_L^2\approx\omega^2$)
where the presence of natural small parameters (such as the smooth
inhomogeneity of the ion density $N(x)$ along the $x$ axis and the
small incidence angles~$\vartheta$ of laser beams at the plasma)
implies the appearance of a hierarchy of components of the
$p$-polarized light wave at the critical plasma point. When
constructing the inherited point RGS, this allows reducing the
total system of six initial equations to a simpler system of two
one-dimensional nonlinear PDEs for the components of the electron
velocity~$v$ and the electric field~$E$ along the density gradient
vector~\cite{KP-KSF89,KP-TMF89},
 \begin{equation}   \label{pl1}
 v_t+a v v_x -  E = 0 ; \quad E_t + a v E_x + \omega_L^2 v = 0 \,,
 \quad \omega_L^2 \equiv  \frac{4\pi \vert e  e_i \vert N}{m} \, .
 \end{equation}
Here, the functions~$v$ and~$E$ are expressed in units of the
dimensionless nonlinearity parameter~$a$, which is proportional to
the value of the magnetic induction~$B$ at the critical point at
the laser frequency; the coordinate~$y$ enters only in combination
with time, $t\to t-(\vartheta y/c)$.

The system of six initial equations admits only a finite group of
point transformations, namely, the group of translations along the
$t$ and $y$ axes for arbitrary $N(x)$. At a constant ion density,
$N=const$, the additional group of $x$-axis translations and the
group of simultaneous rotations in three planes, which are
determined by the coordinates $\{x,y\}$ and the corresponding $x$
and $y$ components of the electron velocity and of the electric
field, arise. In contrast to the initial equations, system
(\ref{pl1})  admits an infinite group of point transformations
with the operator containing the three terms
\begin{equation} \begin{array}{c}
\displaystyle{ X_1 =\mu_1 Y\,; \quad X_2
        =\mu_2 \partial_x + \frac{1}{a}\, Y( \mu_2 ) \partial_v
     +\frac{1}{a}\, Y^2(\mu_2 )\partial_E \,; \ }\\ \mbox{}\\ \displaystyle{
 X_3 =\frac{1}{a}\, \mu_3 \left( a \partial_a - v \partial_v - E \partial_E
    \right) \,; \quad Y = \partial_t + a v \partial_x + E \partial_v -
     \omega_L^2 v \partial_E ,}
\end{array} \label{plgr1}
\end{equation}
each of which contains an arbitrary function $\mu_i$ of
independent and dependent variables and of the parameter~$a$ while
the differential constraints
 $$ Y^3(\mu_2)+Y(\omega_L^2\mu_2)=0\,; \quad Y(\mu_3) = 0 \,.
 \label{plgr2}
 $$
are imposed on $\mu_2$ and $\mu_3$.

To obtain the RGS, point transformation group~(\ref{plgr1})  must
be restricted on a solution of the boundary-value problem that is
approximate over the powers of the parameter~$a$; this solution is
such that the leading approximation for the functions~$v$ and~$E$
is determined by a solution of the linearized system of initial
equations with the corresponding boundary conditions (the
propagation of the electromagnetic wave from the vacuum toward the
plasma) and with the given shape of the density $N(x)$ in the
plasma resonance domain taken into account; the corrections
proportional to $a$ appear when linearizing system~(\ref{pl1}).
Verifying the FS conditions for group~(\ref{plgr1}), we find that
$\mu_1=0$, $\mu_2=-E/\omega^2$, and $\mu_3=1$ for this particular
solution, which gives the desired RGS operator
\begin{equation} \label{plrg1}
R_8= X_2 + X_3 = -\frac{E}{\omega^2}\, \partial_x +  \partial_a \,.
 \end{equation}

The condition of invariance of the solution of the boundary-value problem
w.r.t.\ the RG operator~$R_8$ (the FS condition) is described by the system
of first-order PDEs
 \begin{equation}
 v_a -\frac{E}{\omega^2}\, v_x =0 \,,\quad
 E_a-\frac{E}{\omega^2}\, E_x = 0\,.
 \label{fsspl1} \end{equation}

The solution of the characteristic equations for this system (the system of
Lie equations for the operator $R_8$) reconstructs the PT in the parameter
$a$ up to the exact solution of the boundary-value problem~\cite{KP-TMF89},
 \begin{equation} \begin{array}{c}
 \displaystyle{ E =-\frac{(\omega L)^2}{\Delta} (q_1\sin\omega t+q_2\cos
 \omega t)\,, \quad v=-\frac{\omega L^2}{\Delta}(q_1\cos\omega
 t-q_2\sin\omega t)\,, } \\ \mbox{}\\ \displaystyle{ x=\mu + \varepsilon
 (q_1 \sin\omega t+q_2 \cos\omega t)\,,\quad \varepsilon =aL^2/{\Delta^2}\,,}
 \end{array} \label{solpl1} \end{equation}
which describes the nonlinear structure of the electric field in
the plasma resonance domain. The dimensionless quantities~$x$
and~$\mu$ are normalized to the resonance width~$\Delta$, and $L$
is the characteristic inhomogeneity scale of the plasma ion
density. The choice of the form of the functions $q_1$ and $q_2$
in relations~(\ref{solpl1}) is determined by the actual dependence
on the $x$ coordinate of the electric field, which is a solution
of the initial total system of equations under the corresponding
boundary conditions and with the given shape of the density in the
plasma resonance domain.

For example, for a cold electron plasma with a linear density function, the
functions~$q_1$ and~$q_2$ are
\begin{equation}  \label{solpl2}
 q_1=(1+\mu^2)^{-1}\,, \quad q_2=\mu(1+\mu^2)^{-1}\,,
\end{equation}
where the width $\Delta=(\nu/\omega)L$ is determined by the frequency of
plasma particle collisions.

For a hot plasma in which the heat motion of electrons is
essential, linearizing the initial system of equations, we find
that in the case where the density function is linear in the
plasma resonance domain, the electric field distribution is
expressed through the Airy--Fock functions. The corresponding
nonlinear structures of the electric field and of the density and
velocity of electrons, which appear from using RG
operator~(\ref{plrg1}) to continue the linear relations with the
first nontrivial heat correction (for $a\rightarrow0$) to finite
values of $a$, are determined, as before, by expressions
(\ref{solpl1}) in which the resonance width~$\Delta$ now depends
on the electron temperature, $\Delta=(3V_T^2 L/\omega^2)^{1/3}$,
and the functions $q_i$ are
\begin{equation}   \label{solpl3}
 q_1= \int\limits_{0}^{\infty}d \xi \cos (\mu \xi+ \xi^3/3)\,, \quad
q_2= \int\limits_{0}^{\infty}d \xi \sin (\mu  \xi + \xi^3/3)\,.
\end{equation}
A feature of the formulas for~$v$ and~$E$ in~(\ref{solpl1}) with
the functions $q_1$ and~$q_2$ from~(\ref{solpl3}) is that they
give the exact (at $\omega_L^2= \omega^2$) solution of
Eqs.~(\ref{pl1}), in which the electron pressure is neglected but
the nonzero electron temperature is nevertheless taken into
account.

These formulas as well as their physical consequences were
analyzed in detail in~\cite{KP-TMF89,KP-FP89}. Here, we only note
the value of the results obtained: the FS condition for a simple
mathematical model leads to proper results even in the leading
order in which small corrections to RGSs are neglected (although a
modification of RG operator~(\ref{plrg1}) taking these corrections
into account is not complicated; the corresponding expression for
corrections in the density gradient was given in~\cite{KP-TMF89}).
The big freedom in choosing the functions~$q_1$ and~$q_2$ in
Eqs.~(\ref{solpl1}) permits analyzing the nonlinear structure of
the electric field in the plasma resonance domain in both cold and
hot plasmas uniformly.

We note that RG operator~(\ref{plrg1}) is qualitatively analogous
to Bogoliubov RG operator~(\ref{rg-op}). As in the standard QFT RG
case, operator~(\ref{plrg1}) determines the translation
transformation along a solution characteristic (the parameter~$a$)
as well as more complicated functional transformations of the
coordinate~$x$. The FS property of the physical system under
investigation is the invariance of the desired functions, the
field~$E$ and electron velocity~$v$, w.r.t.\ these RG
transformations. Then, the nonlinearity parameter~$a$, which
enters the RG transformation, is not assumed to be small in
contrast to the terms dropped when obtaining the operator $R_8$ of
the ``approximate" RGS; the smallness of these terms was crucial
for passing from the initial system of equations to
Eqs.~(\ref{pl1}). Therefore, the RGS can be used to extend the
solution obtained in the form of a power series in the
parameter~$a$ to the exact, essentially nonlinear solution,
Eqs.~(\ref{solpl1}). Then, as in the QFT case, the obtained RG
operator corresponds to the exact group transformation w.r.t.\ the
parameter $a$ for a solution of Eqs.~(\ref{pl1}). At the same
time, this RG operator is approximate because it was obtained by
neglecting parameters other than~$a$ and admits a subsequent
improvement w.r.t.\ those parameters.

\subsection{The FS in nonlinear wave optics}

In many cases, we cannot neglect all but leading (zero-order in
small parameters) contributions to an RGS. Therefore, we consider
a variant of the RGS and FS conditions that explicitly takes
contributions of small parameters into account in the RG operator;
such a modification appears when analyzing the boundary-value
problem, which generalizes~(\ref{nlopeq1}) and~(\ref{nlopcond1}),
\begin{equation}  \label{basic}
 \begin{array}{c}
 \displaystyle{ v_{t}+v v_{x} -\alpha n_{x}-\beta\partial_{x}\left(\left
  (x^{1-\nu}/\sqrt{n \,} \right) \partial_{x}\left( x^{\nu-1}\partial_{x}
   \left( \sqrt{n\,}  \right) \right) \right) = 0 \,,  }
\\ \mbox{}\\
\displaystyle{ n_{t} + n v_{x}+ v n_{x} +(\nu-1)\frac{nv}{x} = 0\,.}
\end{array}
\end{equation}
Here, $\alpha$ is the nonlinear refraction parameter as
previously, $\beta$ is the parameter that defines the diffraction
effects, and the respective values $\nu=1$ and $\nu=2$ correspond
to flat and cylindrical geometries of the beam. Boundary
conditions for Eqs.~(\ref{basic})  determine the curvature of the
wave front of the beam and its intensity distribution over the
transverse coordinate~$x$,
\begin{equation} \label{boundary}
 v(0,x) =  V(x) = - x / T \,, \qquad  n(0,x)=N(x) \,.
\end{equation}

\subsubsection{Flat geometry}

We now set the geometry to be flat ($\nu=1$), neglect the
diffraction ($\beta=0$), and consider the simplest case where the
wave beam on the medium boundary (i.e., at $t=0$) has a flat wave
front. Then, Eqs.~(\ref{basic}) become system~(\ref{nlopeq2}). We
write the RGS operators for boundary-value
problem~(\ref{nlopeq2}),~(\ref{nlopcond2}) in canonical
form~(\ref{oprgnl1}). We represent the coordinates~$f$ and~$g$ as
power series,
\begin{equation} \label{coord1}
   f=\sum\limits_{i=0}^{\infty} \alpha^i f^i \,; \quad
   g=\sum\limits_{i=0}^{\infty} \alpha^i g^i\,.
\end{equation}
The method for calculating the coefficients $f^i$ and $g^i$ and
the resulting system of recursive relations for these coefficients
was derived in \cite{K-TMF97,K-TMF99}. There, each coefficient was
determined up to an arbitrary function of $n$, $\chi_{(s)}$, and
of the combination ${\tau}_{(s)}- w(s\chi_{(s)}+n\chi_{(s+1)})$
where $(s)$ denotes the $s$th-order derivative in $n$. This
arbitrariness can be removed by restricting the group on the
desired solution of the boundary-value problem, i.e., by imposing
FS condition (\ref{fssnl1}), which becomes a differential or
algebraic constraint that is compatible with boundary
condition~(\ref{nlopcond2}) at $\tau=0$. Truncating series
(\ref{coord1}), which is possible at small values of~$\alpha$, we
obtain an approximate symmetry; the contributions $f^0$ and $g^0$,
which do not depend on the parameter~$\alpha$, are then determined
by the system of equations
\begin{equation}
 \tau_w - n \chi_n = 0\,, \quad \chi_w  = 0  \,. \label{basic3}
\end{equation}
which is simpler than initial system~(\ref{nlopeq2}). In contrast
to system (\ref{nlopeq2}), which admits only a finite group of
Lie--B\"acklund transformations of a given order,
system~(\ref{basic3}) admits an infinite symmetry group because
the coordinates $f^0$ and $g^0$ can be arbitrary functions of
their arguments. At small~$\alpha$, the symmetry of
Eqs.~(\ref{basic3}) is inherited by initial system~(\ref{nlopeq2})
up to an arbitrary given order in~$\alpha$. Restricting the
obtained approximate group on the solution of the boundary-value
problem, we obtain the desired RGS.

We now turn to RGS operators with coordinates in the form of the binomials
\begin{equation} \label{binom}
f=f^0 + \alpha f^1\,, \qquad g=g^0 + \alpha g^1 \,.
\end{equation}
These operators depend on the functions~$f^0$ and $g^0$, which, in
turn, can be (nonuniquely) determined by the function $N(x)$ of
the transverse intensity distribution at the boundary of the
nonlinear medium.

For the ``soliton" beam intensity distribution function
$N(x)=\cosh^{-2}x$, we can obtain two sets of formulas for the
coefficients resulting from different expressions for~$f^0$
and~$g^0$~\cite{K-TMF97}:
\begin{equation} \label{rgs1_sol} \hspace{-0.4cm}
\begin{array}{l}
\displaystyle{a) \quad f^0 = 2n(1-n)\tau_{nn}- n\tau_n - 2nw(\chi_n +
  n \chi_{nn} )\,, \   f^1 = \frac{1}{2}\,nw^2\tau_{nn}  }\,, \\
\mbox{}\\ \displaystyle{\hphantom{a) \quad} g^0= 2n(1-n)\chi_{nn}
+(2-3n)\chi_n \,,\  g^1= w \left( 2n\tau_{nn} + \tau_n \right)
    +\frac{w^2}{2}\left( n \chi_{nn} + \chi_n \right) \,. }
\end{array}
\end{equation}
\noindent
\begin{equation}  \label{rgs2_sol} \hspace{-2.5 cm}
\begin{array}{l}
\displaystyle{b) \quad f^0 = 1+2n \chi_n \tanh \chi \,, \ f^1=
\left( \frac{\tau^2}{n} - 2 \tau \tau_n + 2 \tau^2 \tanh \chi
\right)\cosh^{-2}\chi} \,, \\
\mbox{}\\ \displaystyle{\hphantom{a)\quad } g^0=0\,, \  g^1= -2
\tau \chi_n \cosh^{-2}\chi -2  \tau_n \tanh \chi \,. }
\end{array}
\end{equation}
Calculating contributions of higher
orders~\cite{K-TMF97,KPS-DE93,KP-MCM97}, we find that the
functions~$f^i$ and $g^i$ vanish for $i\geq2$ in case~{\sl a} and
formulas (\ref{rgs1_sol})  and~(\ref{coord1}) describe an exact
RGS; the comparison with coordinates~(\ref{rgs1-sol}) demonstrates
that the functions~$f^{1,2}$ and $g^{1,2}$ are the coefficients of
the coordinate expansions in powers of $\alpha$. In case~{\sl b},
series~(\ref{coord1}) do not terminate, and formulas
(\ref{rgs2_sol}) and~(\ref{coord1}) pertain to an approximate RGS.

For the wave beam with a Gaussian initial intensity shape, $N(x)=\exp(-x^2)$,
we have two sets of the $f$ and $g$ operators of an approximate RGS:
 \begin{equation}
 \label{rgs1_gauss} \hspace*{-1.43cm}
 \begin{array}{l}
 \displaystyle{a) \ f^0 = 1+ 2n\chi \chi_n \,,  \quad g^0=0\,,
 \quad f^1 = -2 \tau \tau_n +\frac{\tau^2}{n} \,, \quad g^1= -2
 \left(\tau \chi_n + \chi \tau_n  \right) \,,}
 \end{array}
 \end{equation}
 \begin{equation} \label{rgs2_gauss}
 \hspace*{-0.52cm}
 \begin{array}{l} \displaystyle{b)\ f^0 =  2n(\tau \chi_n
 + \tau_n\chi) \,,\ \ g^0 = 1 + 2n\chi \chi_n\,,} \ \
 \displaystyle{ f^1 = 2 \chi \tau_{\alpha} \,,\ \ g^1= 2 \left(\chi
 \chi_{\alpha} - \tau \tau_n  \right) \,.}
 \end{array}
 \end{equation}
Formulas~(\ref{rgs1_sol})--(\ref{rgs2_gauss})  demonstrate the
main advantage of the approximate RGS method, which permits
analyzing boundary-value problems with arbitrary boundary data,
which are expressed through differential or algebraic expressions
for the functions~$f^0$ and $g^0$. The example of
operator~(\ref{rgs1_sol}) shows that in some cases the approximate
RGS method can result in an exact RGS whose presence must be
established using the tools in the previous section. Because
expressions~(\ref{rgs1_sol}) for the coordinates $f^i$ and $g^i$
contain second-order derivatives, the corresponding operator $R$
is the RGS Lie--B\"acklund operator of the second order, while
operators~(\ref{rgs2_sol})--(\ref{rgs2_gauss}) are equivalent to
point symmetry operators. Meanwhile, because
operator~(\ref{rgs2_gauss}) contains the first derivative in the
parameter~$\alpha$, the FS transformations also involve the
nonlinear refraction parameter~\cite{K-TMF97}.
(See~\cite{K-TMF97,KSh-JNOPM97} for a detailed analysis of
formulas~(\ref{rgs1_sol})--(\ref{rgs2_gauss}) and the physical
consequences of their substitution in FS
conditions~(\ref{fssnl1}).)

\subsubsection{Cylindrical case}

Analogously to the previous case, approximate RGSs for the case
$\beta\neq0$ can be constructed using FS
conditions~\cite{K-TMF99}. An example is the RGS operator for the
{\sl cylindrical} $(\nu=2)$ wave beam,
\begin{equation}  \label{rgsym}
\begin{array}{l}
\displaystyle{ R_9 = \left[  \left(  1-\frac{t}{T} \right)^2 + t^2
S_{\chi\chi} \right]    \partial_{t}
 + \left[- \frac{x}{T} \left( 1 - \frac{t}{T} \right)
   + t S_{\chi} + v t^2 S_{\chi\chi} \right]    \partial_{x} }
  \\   \mbox{}  \\
\displaystyle{+\left[\frac{x}{T^2} +\frac{v}{T}\left(1-\frac{t}{T} \right)
    + S_{\chi} \right]   \partial_{v}
   + \left[ \frac{2n}{T} \left( 1 - \frac{t}{T} \right)
   - nt\left(1+\frac{vt}{x} \right) S_{\chi\chi}
   - \frac{nt}{x} S_{\chi} \right] \partial_{n} \,.}
\end{array}
\end{equation}
where the function~$S$ depends on the variable $\chi=x-vt$,
$$
 S(\chi) = \alpha N(\chi) + \frac{\beta}{\chi\sqrt{N(\chi)}} \,
 \partial_{\chi} \left( \chi \partial_{\chi} \left( \sqrt{N(\chi)}\right)
 \right) \, ,  $$
contains two small parameters~$\alpha$ and~$\beta$, and is
determined by the initial beam intensity distribution function~$N$
at the boundary of the nonlinear medium. Then, as in the case
$\beta=0$, there exist such functions $N$ for which
operator~(\ref{rgsym}) corresponds to an exact, rather than
approximate, RGS, i.e., the RGS that is applicable at arbitrary,
not necessarily small, values of the parameters~$\alpha$
and~$\beta$~\cite{K-TMF99}. The FS transformations appear from the
solution of the characteristic equations for the first-order PDE
that is conjugate to~(\ref{rgsym}) and permit continuing the PT
solutions, which are determined only in a small vicinity of the
nonlinear medium boundary, to a domain where essentially nonlinear
effects prevail~\cite{K-TMF99}.

To conclude this section, we again note that we can write the FS
conditions in many different ways, which implies different
approaches to their subsequent use in problems with a small
parameter. The use of approximate symmetries results in some
peculiarities in constructing the RGSs and in using the FS
conditions.

First, as discussed in the previous section, we apply the FS
conditions in the form of a system of higher-order DEs and analyze
them together with the initial equations. In contrast to finite FS
transformations, which are infinite formal series
(see~\cite{Sprav}, Vol.~3, Chap.~1), the differential formulation
of the FS conditions is suitable for constructing solutions of a
boundary-value problem.

Second, the FS conditions expressed as approximate symmetries are
additional differential constraints, which together with the
initial equations determine the RG manifold. In turn, such a
manifold can be used to construct approximate point RGSs.

Third, the FS conditions are important at all stages of
constructing approximate RGSs already starting with setting the
RGS operator coordinates for an unperturbed solution. New
prospects are provided by the possibility of constructing RGSs
based on approximate symmetries for problems with arbitrary
boundary conditions.

Fourth, we introduce parameters that are used to construct approximate RGSs
in the set of variables of the FS transformations.

\section{Conclusion} \label{sec6}

We now formulate how the FS notion has been transformed in the
last decade. This notion first appeared as pertaining to the group
transformations developed by Lie in~\cite{Slie}, which was used to
construct the QFT (Bogoliubov) RG~\cite{stp,bsh55}. The latter
concept was based on the one-parameter Lie group of local
transformations, the class the Bogoliubov RG belongs to. The FS
notion unified various RGs and exhibited their intrinsic group
structure.

The progress in applying mathematical methods to the investigation
of symmetries of differential and integral-differential equations,
which resulted in modern group analysis (see,
e.g.,~\cite{Oves,Sprav,Olver,Vinogradov} and references therein),
has led to the transformation of the FS notion.

First, this resulted in constructing the special class of
symmetries related to the RG transformations, namely, the RGSs.
Such a method (see Sec.~2) regularizes the procedure for finding
RG transformations, at least for systems that are described by
differential and integral-differential equations. Using the
infinitesimal approach, we can formulate the RG invariance
conditions through the RGS operators and can therefore use all the
powerful tools of modern group analysis to construct these
conditions. As a result, the RGS notion now includes not only
point symmetries (Sec.~3) but also Lie--B\"acklund symmetries
(higher, or generalized, symmetries) (Secs.~4 and~5), approximate
symmetries (Sec.~5), etc. The list of examples, which is far from
complete, presented in this paper includes the most advanced
results in constructing the RGSs; furthermore, our approach is
equally applicable to the cases of nonlocal RGSs and to RGSs of
integral-differential equations if, for example, we apply the
method described in~\cite{KKP93}. Therefore, modern group analysis
is as important for developing the FS notion as was the classical
Lie group analysis for establishing the Bogoliubov RG.

The FS conditions are formulated as conditions of invariance
w.r.t.\ transformations determined by RG operators. This becomes
important when the formulation of finite FS transformations is
doubtful, e.g., for the Lie--B\"acklund RGSs (see Secs.~4 and~5).
Being universal, this formulation is useful not only at the stage
of finding the solutions but also when constructing RGS operators.
Thus, the FS conditions and the procedure for their verification
became important components of the RGS construction. The
conditions themselves can be algebraic as well as differential
relations containing derivatives of higher (not necessarily first,
see Sec.~4.2) order.

A feature of the new way of constructing RGSs is the regular
procedure for constructing Lie algebras of finite dimension.
Numerous examples of such algebras are given
in~\cite{KPSh-JMP98,KP-LGA94}, although they have not yet found
wide practical applications (see Sec.~3). Such Lie algebras
correspond to {\sl multiparameter groups}. The RGS apparatus
results in a set of group operators constituting a Lie algebra.
Such groups are customarily used by researchers into symmetries of
equations arising in contemporary theoretical
physics.\footnote{For instance, the basis of the Poincar\'e group
transformation algebra consists of 10 generators.}\ However, the
``renormalization group" in both the QFT (Bogoliubov) and the
Wilson approaches was always assumed to be a one-parameter group.
Rare attempts to consider two-parameter constructions have always
(to the best of our knowledge~\cite{kris}) led to direct products
of two one-parameter RGs.

The RG technique was customarily used in problems with
singularities to improve the PT and to give a correct description
of the solution behavior in the vicinity of a singularity; these
properties are intrinsic in our approach as well. Namely, using
the FS conditions formulated in Secs.~4 and~5 on the base of the
Lie--B\"acklund RGS, we were able not only to describe the
structure of the known singular solutions but also to find new
ones (see \cite{KSh-JNOPM97,K-TMF97,K-TMF99}). We stress that the
RGS approach to these problems results in a structure of
two-dimensional singularities that differs from the structure of
singularities appearing within the Bogoliubov RG setting.

\section{Acknowledgments}

This work was supported in part by the Russian Foundation for Basic Research
(Grant Nos.~96-15-96030 and~99-01-00232) and INTAS (Grant No.~96-0842).

 \newpage
\tableofcontents
\end{document}